\documentclass[prl,twocolumn,showpacs]{revtex4}
\usepackage{amsfonts,amsmath}
\newcommand{\blambda}{\boldsymbol \lambda}

\begin{document}

\title  {Statistics of Real Eigenvalues in Ginibre's Ensemble of Random Real
Matrices}

\author {Eugene Kanzieper$\,{}^\flat$ and Gernot Akemann$\,{}^\sharp$}

\affiliation
        {${}^\flat$Department of Applied Mathematics, Holon Academic Institute of Technology, Holon 58102, Israel\\
         ${}^{\sharp}$Department of Mathematical Sciences, Brunel
         University West London, Uxbridge UB8 3PH, United Kingdom
        }
\date   {21 July 2005}
\begin  {abstract}
The integrable structure of Ginibre's orthogonal ensemble of random
matrices is looked at through the prism of the probability $p_{n,k}$
to find exactly $k$ real eigenvalues in the spectrum of an $n\times
n$ real asymmetric Gaussian random matrix. The exact solution for
the probability function $p_{n,k}$ is presented, and its remarkable
connection to the theory of symmetric functions is revealed. An
extension of the Dyson integration theorem is a key ingredient of
the theory presented.
\end{abstract}

\pacs{02.50.-r, 05.40.-a,
75.10.Nr\;\;\;\;\;\;\;\;\;\;\;\;\;\;\;\;\;\;\;\;\;\;\;\;\;\;\;\;\;\;\;\;\;\;\;\;\;\;\;\;\;
DOI:\,10.1103/PhysRevLett.95.230201} \maketitle

{\it Introduction}.---In the mid-1960's, Ginibre introduced
\cite{G-1965} statistical ensembles of (i) real, (ii) complex, and
(iii) real quaternion random matrices whose eigenvalues may belong
to any point of the complex plane ${\mathbb C}$. They were derived
from the celebrated Gaussian orthogonal (GOE), Gaussian unitary
(GUE), and Gaussian symplectic (GSE) random matrix ensembles
\cite{M-2004} in a purely formal way by dropping the Hermiticity
constraint and, therefore, can be thought of as their direct
non-Hermitean descendants. Respectively coined as GinOE, GinUE, and
GinSE, non-Hermitean random matrix models exhibited intriguingly
rich mathematical structures whose complexity exceeded by far that
of Hermitean random matrix theory (RMT).

From the physical point of view, non-Hermitean random matrices have
proven to be as important \cite{FS-2003} as their Hermitean
counterparts \cite{GGW-1998}. Ginibre's random matrices appear in
the description of dissipative quantum maps \cite{GHS-1988-89}
(GinUE), in the studies of dynamics \cite{SCSS-1988} and the
synchronisation effect \cite{TWG-2002-04} in random networks
(GinOE), in the statistical analysis of cross-hemisphere correlation
matrix of the cortical electric activity \cite{KDI-2000} (GinOE),
and in the characterisation of two-dimensional random space-filling
cellular structures \cite{LH-1990-93} (GinUE). They also arise in
the context of ``directed quantum chaos'' \cite{E-1997,FKS-1997}
(GinOE, GinUE, GinSE). Chiral deformations of non-Hermitean random
matrices (GinOE, GinUE, GinSE) help elucidate universal aspects of
the phenomenon of spontaneous chiral symmetry breaking in quantum
chromodynamics \cite{QCD}: the presence or absence of real
eigenvalues singles out different breaking patterns. More recent
findings \cite{Z-2002} associate statistical models of non-Hermitean
normal random matrices with integrable structures of conformal maps
and interface dynamics at both classical \cite{MWZ-2000} and quantum
scales \cite{ABWZ-2002}. For a comprehensive review of these and
other physical applications, the reader is referred to Ref.
\cite{FS-2003}.

Out of the three non-Hermitean matrix models\cite{G-1965}, Ginibre's
orthogonal ensemble defined by the probability density $   P_{{\rm
GinOE}}[{\cal H}] = (2\pi)^{-n^2/2} \exp
    \left[ - \, {\rm tr\,} \left( {\cal H} {\cal
    H}^{\rm \,T}/2 \right)\right]
$ for an $n \times n$ real matrix ${\cal H}$ to occur is the least
studied and the most challenging. Perhaps, the great difficulties
faced in statistical analysis of the GinOE can be attributed to the
fact that its, generically complex, spectrum $(w_1,\cdots, w_n)$ may
contain a {\it finite fraction} of real eigenvalues. This very
peculiar feature of GinOE can conveniently be accommodated by
dividing the entire space ${\mathbb T}(n)$ spanned by all real
$n\times n$ matrices ${\cal H}\in {\mathbb T}(n)$ into $(n+1)$
mutually exclusive sectors ${\mathbb T}(n/k)$ associated with the
matrices having exactly $k$ real eigenvalues, such that ${\mathbb
T}(n)=\bigcup_{k=0}^n {\mathbb T}(n/k)$. The sectors ${\mathbb
T}(n/k)$, characterised by the partial probability densities
$P_{{\cal H}\in {\mathbb T}(n/k)} (w_1,\cdots, w_n)$, can be
explored separately because they contribute additively to the joint
probability density function (JPDF) of all $n$ eigenvalues of ${\cal
H}$ from ${\mathbb T}(n)$:
\begin{equation}
    P_n(w_1,\cdots, w_n) =
    \sum_{k=0}^n P_{{\cal H}\in {\mathbb T}(n/k)}
    (w_1,\cdots, w_n).
\label{jpdf=1}
\end{equation}
Two spectral characteristics are of particular physical interest:
(i) the partial $(r_1,r_2)$-point correlation functions obtained by
integrating out all but $r_1$ real and $r_2 $ complex eigenvalues in
$P_{{\cal H}\in {\mathbb T}(n/k)}$, and (ii) the probability
\begin{equation}
\label{pnk=int}
    p_{n,k} = \prod_{i=1}^k \int_{{\mathbb R}} d\lambda_i
    \prod_{j=1}^\ell \int_{{\mathbb R}} d{\rm Re\,}z_j
    \int_{{\mathbb R^+}} d{\rm Im\,}z_j\,
P_{{\cal H}\in {\mathbb T}(n/k)}
\end{equation}
to find exactly $k$ real eigenvalues in the spectrum of \linebreak
GinOE that additionally contains $\ell$ pairs of complex conjugated
eigenvalues, so that $n=k+2\ell$. {\it The probability function
$p_{n,k}$ is the central object of our study}, which, if considered
in a wider context, aims to highlight an exclusive r\^ole played by
symmetric functions in the description of GinOE.

The most important and general result available for GinOE is due to
the breakthrough work by Lehmann and Sommers \cite{LS-1991} who
proved, a quarter of a century after Ginibre's work, that the $k$-th
partial JPDF ($0\le k \le n$) equals \cite{RP}
\begin  {widetext}
\begin{eqnarray} \label{b=1a}
    P_{{\cal H}\in {\mathbb T}(n/k)}
    =
    \frac{2^{\ell - n(n+1)/4}}{i^\ell \,k! \,\ell!\, \prod_{j=1}^n \Gamma(j/2)}
    \prod_{i>j=1}^k |\lambda_i - \lambda_j|
    \, \prod_{j=1}^k \prod_{i=1}^\ell (\lambda_j-z_i)(\lambda_j-{\bar
    z}_i) \,
    \prod_{j=1}^k \, e^{-\lambda_j^2/2}
    \prod_{i>j=1}^{\ell} |z_i - z_j|^2 |z_i - {\bar
    z}_j|^2\,
    \nonumber
\end{eqnarray}
\end{widetext}
\begin{eqnarray} \label{b=1}
    \times \,
    \prod_{j=1}^\ell \, (z_j-{\bar z}_j)\,
     {\rm erfc} \left(
        \frac{z_j-{\bar z}_j}{i\sqrt{2}}
    \right)\,
    e^{-(z_j^2+{\bar z}_j^2)/2}.
\end{eqnarray}
In writing (\ref{b=1}), we have used a representation due to Edelman
\cite{Ed-1997} who rediscovered the result \cite{LS-1991} a few
years later. The above JPDF is supported for
$(\lambda_1,\cdots,\lambda_k)\in {\mathbb R}^k$, $({\rm Re}\,
z_1,\cdots,{\rm Re}\, z_\ell)\in {\mathbb R}^\ell$, and $({\rm Im}\,
z_1,\cdots, {\rm Im}\, z_\ell)\in ({\mathbb R}^+)^\ell$. A
particular case $k=n$ of (\ref{b=1}), corresponding to the matrices
${\cal H} \in {\mathbb T}(n/n)$ with all eigenvalues real, was first
derived by Ginibre \cite{G-1965}.

Equations (\ref{jpdf=1}) and (\ref{b=1}) solve the problem of
finding the JPDF of all $n$ eigenvalues in GinOE. Despite this
tremendous progress, a calculation of the $(r_1,r_2)$-point
correlation function based on (\ref{b=1}) turns out to be {\it the}
problem, particularly, because the well-developed machinery
\cite{M-2004} of RMT, which was at use in studies of complex
eigenvalue correlations in GinUE \cite{G-1965} and GinSE
\cite{MS-1966,K-2002}, fails to work in this case. In particular,
the famous Dyson integration theorem \cite{D-1970} (Theorem 5.1.4 in
Ref. \cite{M-2004}) is no longer applicable to the study of spectral
statistics in GinOE, as will be reasoned below. This failure clearly
signals of novel mathematical structures lurking behind (\ref{b=1}).

To unveil and explore these structures, we wish to concentrate on
the probability function $p_{n,k}$. Previous attempts by Edelman and
co-workers \cite{Ed-1997,EKS-1994} to attack the problem brought no
explicit formula for $p_{n,k}$ for generic $k$. The analytic results
currently available include: (i) the probability of having all $n$
eigenvalues real equals \cite{Ed-1997} $p_{n,n}=2^{-n(n-1)/4}$ (this
is the smallest probability out of all $p_{n,k}$); (ii) for all
$0\le k\le n$, the $p_{n,k}$ is of the form \cite{Ed-1997}
$p_{n,k}=r_{n,k}+ s_{n,k}\sqrt{2}$, where $r_{n,k}$ and $s_{n,k}$
are rational; (iii) the expected number $E_n=\sum_{k=0}^n
k\,p_{n,k}$ of real eigenvalues is given by \cite{EKS-1994}
\begin{eqnarray}
\label{E-n}
    E_n =\frac{1}{2} + \sqrt{2} \,\, \frac{{}_2 F_1 (1,-1/2;n;
1/2)}{B(n,1/2)}.
\end{eqnarray}

{\it Main results}.---The following four statements summarise the
new results reported in this Letter.

{\it Statement 1}. The probability $p_{n,k}$ of exactly $k$ real
eigenvalues occurring equals
\begin{eqnarray}
\label{pnk=sol}
    p_{n,k}=p_{n,n-2\ell}= p_{n,n}\,\, {\cal F}_\ell (p_1,\cdots,
    p_\ell).
\end{eqnarray}
The universal multivariate polynomials
\begin{eqnarray}
\label{FL}
    {\cal F}_\ell(p_1,\cdots,p_\ell) =
    (-1)^\ell \sum_{|\blambda|=\ell} \, \prod_{j=1}^g \frac{1}{\sigma_j!}
\left(
    - \frac{p_{\ell_j}}{\ell_j}
\right)^{\sigma_j}
\end{eqnarray}
\begin{table}[b]
\begin{tabular}{c| c l  r |} \hline\hline
$\;k\;$ & Exact Solution& {\;\;Appr.}  & {Numerics$\;$} \\
\hline\hline
${}$ & ${}$ & ${}$ & ${}$ \\
$0$ & {\large $\,\frac{29930323227453 -
    20772686238032\sqrt{2}}{17592186044416}\,$} & $0.031452$ & $0.031683\;$
\\
${}$ & ${}$ & ${}$ & ${}$ \\
 \hline
${}$ & ${}$ & ${}$ & ${}$\\
$2$  & {\large $\,\frac{3(1899624551312\sqrt{2} -
2060941421503)}{4398046511104}$} & $0.426689$ & $0.427670\;$
\\
${}$ & ${}$ & ${}$ & ${}$ \\
\hline
${}$ & ${}$ & ${}$ & ${}$ \\
$4$ & {\large $\,\frac{3(2079282320189 -
505722262348\sqrt{2})}{8796093022208}$} & $0.465235$ & $0.464098\;$ \\
${}$ & ${}$ & ${}$ & ${}$ \\
\hline
${}$ & ${}$ & ${}$ & ${}$ \\
$6$ & {\large $\,\frac{252911550974\sqrt{2}-27511352125}{4398046511104}$} & $0.075070$ & $0.075021\;$ \\
${}$ & ${}$ & ${}$ & ${}$ \\
\hline
${}$ & ${}$ & ${}$ & ${}$ \\
$8$ & {\large
$\,\frac{15(1834091507-10083960\sqrt{2})}{17592186044416}$} &
$0.001552$ & $0.001526\;$ \\
${}$ & ${}$ & ${}$ & ${}$ \\
\hline
${}$ & ${}$ & ${}$ & ${}$ \\
$10$ & {\large $\,\frac{3(1260495\sqrt{2}-512)}{2199023255552}$} &
$0.000002$ & $0.000002\;$\\
${}$ & ${}$ & ${}$ & ${}$ \\
\hline
${}$ & ${}$ & ${}$ & ${}$ \\
$12$ & {\large $\,\frac{1}{8589934592}$} & $0.000000$ & $0.000000\;$ \\
${}$ & ${}$ & ${}$ & ${}$ \\
\hline\hline
\end{tabular}
\caption{Exact solution for $p_{12,k}$ (second and third column)
compared to numeric simulations (fourth column) performed by direct
diagonalisation of $1,000,000$ of $12\times 12$ matrices. }
\label{PNK}
\end{table}are exclusively determined  by the partitions $\blambda$ of the size
$|\blambda|=\ell$, where $\ell$ is the number of pairs of
complex-conjugated eigenvalues. The formula (\ref{FL}) uses a
frequency representation \cite{R-FR} of those partitions \linebreak
$\blambda = (\ell_1^{\sigma_1},\cdots, \ell_g^{\sigma_g})$. Experts
in the theory of symmetric functions may readily recognise that our
${\cal F}_\ell$'s are, up to a factorial, so-called {\it zonal
polynomials} \cite{M-1998}
\begin{eqnarray}
\label{FL-Zon} {\cal F}_\ell(p_1,\cdots,p_\ell) = \frac{1}{\ell!}\,
Z_{(1^\ell)} (p_1,\cdots,p_\ell).
\end{eqnarray}
They are tabulated in the manuscript \cite{J-1976-2005} by Jack and
can also be efficiently calculated \cite{M-1998} by recursion. On
the contrary, the arguments $p_j$'s of the multivariate polynomials
in (\ref{pnk=sol}) are matrix model dependent:
\begin{eqnarray} \label{traces}
    p_j = {\rm tr}_{(0,[n/2]-1)} \, {{\hat \varrho}\,}^j.
\end{eqnarray}
(The notation $[x]$ stands for the integer part of $x$). The
non-universal matrix ${\hat \varrho}$ is sensitive to the parity of
$n$. For $n=2m$ even, its entries are
\begin{eqnarray}
    \label{rho-even}
    \hspace{-0.2cm}
    {\hat \varrho}_{\alpha,\, \beta}^{\rm even} &=&
    \int_0^\infty dy\, y^{2(\beta-\alpha)-1} \, e^{y^2}\, {\rm
    erfc}(y \sqrt{2})
    \Big[
        (2\alpha+1) \,\nonumber \\
    &\times& L_{2\alpha+1}^{2(\beta-\alpha)-1}(-2y^2)
        + 2y^2 \,
        L_{2\alpha-1}^{2(\beta-\alpha)+1}(-2y^2)
    \Big], 
\end{eqnarray}
while for $n=2m+1$ odd,
\begin{eqnarray}
   \label{rho-odd}
    {\hat \varrho}_{\alpha,\,\beta}^{\rm odd} =
    {\hat \varrho}_{\alpha,\,\beta}^{\rm even} - (-4)^{m-\beta}
    \frac{m!}{(2m)!}\frac{(2\beta)!}{\beta!}\,
    {\hat \varrho}_{\alpha,\,m}^{\rm even}.
\end{eqnarray}
Here, $L_j^\alpha(x)$ are generalised Laguerre polynomials.

We wish to stress that zonal polynomials $Z_{(1^\ell)}$ are the
integral part of the {\it final solution} (\ref{pnk=sol}), and not
merely the means of its derivation. This highlights a strong
connection between integrable structure of GinOE, integer partitions
(\ref{FL}), and the theory of symmetric functions (\ref{FL-Zon}).

{\it Statement 2}. The entire generating function for the
probabilities $p_{n,n-2\ell}$ can be reconstructed from
(\ref{pnk=sol}). The fairly compact result
\begin{equation}
    \label{gf=1}
    G_n(z) = \sum_{\ell=0}^{[n/2]} z^\ell p_{n,n-2\ell} = p_{n,n} \, \det
    (\openone_{[n/2]} + z\, {\hat \varrho})
\end{equation}
with the ${\hat \varrho}$ of needed parity provides us with yet
another way of computing the entire set of $p_{n,k}$'s at once. For
comparison of our analytic predictions with numeric simulations, see
the Table I.

{\it Statement 3}. The previously unknown JPDF of $\ell$ pairs of
complex conjugated eigenvalues of a matrix ${\cal H}\in {\mathbb
T}(n/k)$ is given by the formula (\ref{jpdf-c=1}).

{\it Statement 4}. This statement is formulated in the form of the
Pfaffian integration theorem (\ref{theorem-b}) below. (It is highly
likely that this theorem will have implications far beyond the scope
of the present Letter).

{\it Sketch of the derivation.}---Because of the space limitations,
only key points of the derivation will be presented below; the
technical details will be reported elsewhere \cite{KA}.

To determine the probability function $p_{n,k}$, defined by
(\ref{pnk=int}), we first integrate out all
$(\lambda_1,\cdots,\lambda_k)$ in (\ref{b=1}) which can be viewed as
a {\it hybrid} of GOE \cite{M-2004} and GinSE \cite{MS-1966,K-2002}.
The most efficient way to do so is to spot that the first line in
(\ref{b=1}), if taken without any prefactors therein, is
proportional to the average characteristic polynomial \linebreak
$\langle \prod_{j=1}^\ell \det(z_j - H) \det({\bar z}_j - H)\rangle$
of a $k\times k$ random matrix $H$ drawn from the GOE with the
weight function $\exp(-{\rm tr\,} H^2/2)$. It admits the Pfaffian
representation \cite{NN-2001,BS-2004}
\begin{eqnarray}
    \label{CP}
    \frac{k!}{n!}\,
    \, \frac{c_n/c_k}{\Delta_{2\ell}(\{z,{\bar z}\})}\,
    {\rm pf\,} \left[
    \begin{array}{cc}
      K_{n}(z_i, z_j) & K_{n}(z_i, {\bar z}_j) \\
      K_{n}({\bar z}_i, z_j) & K_{n}({\bar z}_i, {\bar z}_j) \\
    \end{array}
    \right]_{2\ell \times 2\ell}
\end{eqnarray}
with the constant $c_k= 2^{k/2} k!\,\prod_{j=1}^k \Gamma(j/2)$ given
by a GOE Selberg's integral \cite{M-2004}. The Vandermonde
determinant $\Delta_{2\ell}(\{z,{\bar z}\})$ is calculated on the
set of complex eigenvalues $(z_1,\cdots,z_\ell; {\bar z}_1,\cdots,
{\bar z}_\ell)$. The kernel function $K_{n}(z,z^\prime)$ in
(\ref{CP}) is a so-called $D$-part \cite{TW-1998} of the GOE matrix
kernel \cite{M-2004} that can be expressed \cite{K-rem} in terms of
arbitrary monic polynomials $q_j$:
\begin{eqnarray}
\label{Dker}
    K_n(x,y) = \frac{1}{2} \sum_{j,k=0}^{n-1} q_j(x) \, {\hat \mu}_{jk}\,
    q_k(y).
\end{eqnarray}
The real antisymmetric matrix ${\hat \mu}$ is determined by its
inverse
\begin{equation}
\label{mu}
    ({\hat \mu}^{-1})_{jk} = \frac{1}{2} \int_{{\mathbb R}^2} dx\,dy\,
    e^{-(x^2+y^2)/2} {\rm sgn}(y-x) \, q_j(x)\, q_k(y).
\end{equation}
Substituting (\ref{CP}) into (\ref{pnk=int}) completes the
$\lambda$-integration therein, bringing a new result for the JPDF of
$\ell$ pairs of complex conjugated eigenvalues of ${\cal H}\in
{\mathbb T}(n/k)$:
\begin{widetext}
\begin{eqnarray}
    \label{jpdf-c=1}
    P_{{\cal H}\in {\mathbb T}(n/k)} (z_1,{\bar z}_1,\cdots, z_\ell,{\bar
    z}_\ell) =  \frac{p_{n,n}}{\ell !}\,
    \left( \frac{2}{i} \right)^\ell
    \prod_{j=1}^\ell \, {\rm erfc}
    \left( \frac{z_j - {\bar z}_j}{i\sqrt{2} } \right) \, e^{
        - (z_j^2 + {\bar z}_j^2)/2}\,
    {\rm pf\,} \left[
    \begin{array}{cc}
      K_{n}(z_i, z_j) & K_{n}(z_i, {\bar z}_j) \\
      K_{n}({\bar z}_i, z_j) & K_{n}({\bar z}_i, {\bar z}_j) \\
    \end{array}
    \right]_{2\ell \times 2\ell}.
\end{eqnarray}
The structure of (\ref{jpdf-c=1}) mirrors that of the JPDF of all
complex eigenvalues in the GinSE (see, e.g., Chap. 15.2 in Ref.
\cite{M-2004}, and Ref. \cite{K-2002}), possibly triggering one to
think that the remaining $z$-integrations in (\ref{pnk=int}) could
readily be accomplished by virtue of the Dyson integration theorem
\cite{D-1970,M-2004}. Sadly, this is not the case because the
$D$-part $K_n(z,z^\prime)$ of the GOE matrix kernel does {\it not}
obey the projection property in the {\it complex plane}
\begin{eqnarray}
    \label{no-proj}
    \int_{{\rm Im\,}w>0}\,d\alpha(w)\, K_n(z,{\bar w})\, K_n(w, z^\prime)
    \neq \frac{1}{2} K_n(z,z^\prime)
\end{eqnarray}
with respect to the measure $d\alpha(w)={\rm erfc}[(w-{\bar
w})/i\sqrt{2}]\exp[-(w^2+{\bar w}^2)/2]\, d^2w$ from
(\ref{jpdf-c=1}) (compare to the Lemma 15.2.1 of Ref.
\cite{M-2004}). Fortunately, the projection property is not actually
necessary to carry out the integrations. We were able to prove the
following Pfaffian integration theorem \cite{KA}.

{\it Theorem.}---Let $d\alpha(z)$ be any benign measure on $z \in
{\mathbb C}$, and the kernel function $K_n(x,y)$ be an antisymmetric
function specified by (\ref{Dker}). Then, the integration formula
\begin{eqnarray}
    \label{theorem-b}
    \;\;\;
    \left(\frac{2}{i}\right)^\ell \prod_{j=1}^\ell \, \int_{z_j\in {\mathbb C}} d\alpha(z_j)\,\,
    {\rm pf\,} \left[
    \begin{array}{cc}
      K_{n}(z_i, z_j) & K_{n}(z_i, {\bar z}_j) \\
      K_{n}({\bar z}_i, z_j) & K_{n}({\bar z}_i, {\bar z}_j) \\
    \end{array}
    \right]_{2\ell \times 2\ell}
    =
    Z_{(1^\ell)}\left(
    \frac{1}{2} {\rm tr}_{(0,n-1)}
    {\hat \sigma}^1,\cdots,\frac{1}{2} {\rm tr}_{(0,n-1)}
    {\hat \sigma}^\ell
    \right)
\end{eqnarray}
\end{widetext}
holds, provided the integrals in the l.h.s. exist.

The Pfaffian integration formula contains particular zonal
polynomials \cite{M-1998} $Z_{(1^\ell)}$ whose arguments are traces
of ${\hat \sigma}^j$ with
\begin{equation}
    \label{sigma=def}
    {\hat \sigma}_{\alpha,\beta} = i \sum_{k=0}^{n-1} {\hat \mu}_{\alpha,k}
    \int d\alpha(z) \left[
        q_k(z)\, q_\beta({\bar z}) - q_\beta(z) \, q_k({\bar z})
    \right].
\end{equation}
Operationally, the proof \cite{KA} is based on a topological
interpretation of the permutational expansion of Pfaffian in
(\ref{theorem-b}) combined with term-by-term integration.

Equipped with this theorem, we are ready to complete the
$z$-integrations in (\ref{pnk=int}). The calculation is most
economic in the basis where monic polynomials $q_j$'s coincide with
GOE skew-orthogonal polynomials. For $n=2m$ even, they are given by
\cite{M-2004}
\begin{eqnarray}
\label{ops-1}
    q_{2j+1}(x) &=& 2^{-(2j+1)} H_{2j+1}(x) - 2^{-(2j-1)}j\,
    H_{2j-1}(x), \nonumber \\
    q_{2j}(x) &=& 2^{-2j} H_{2j}(x),
\end{eqnarray}
so that
\begin{equation}
\label{ops-2}
    {\hat \mu}_{jk} = \frac{1+(-1)^k}{2^{1-k}k!\,\sqrt{\pi}} \, \delta_{j, k+1}  -
     \frac{1+(-1)^j}{2^{1-j}j!\,\sqrt{\pi}} \, \delta_{j, k-1}.
\end{equation}
Substituting (\ref{ops-1}) and (\ref{ops-2}) into (\ref{sigma=def}),
and taking into account (\ref{theorem-b}), we reproduce the
announced solution (\ref{pnk=sol}) -- (\ref{rho-even}) after a bit
lengthy but straightforward calculations. The case $n=2m+1$ odd can
be treated similarly \cite{KA} leading to the same solution but with
(\ref{rho-even}) replaced by (\ref{rho-odd}).

It remains to establish the result (\ref{gf=1}) for the generating
function $G_n(z)$. It stems from the summation formula
\begin{eqnarray}
\label{sf}
    \sum_{r=0}^\infty \frac{ z^r}{r!} \, Z_{(1^r)}(p_1,\cdots,p_r) =\exp
    \Bigg(
        \sum_{r\ge 1} (-1)^{r-1} \frac{p_r z^r}{ r }
        \Bigg)
\end{eqnarray}
well known in the theory of symmetric functions \cite{M-1998}.
Identifying $p_r = {\rm tr}_{(0,[n/2]-1)}{\hat \varrho}^r$, the
exponent in (\ref{sf}) can be evaluated explicitly to confirm
(\ref{gf=1}).

{\it Conclusion.}---To summarise, the exact solution was presented
for the probability $p_{n,k}$ to find precisely $k$ real eigenvalues
in the spectrum of an $n\times n$ random matrix drawn from GinOE.
Expressed in terms of zonal polynomials, the solution associates the
integrable structure of GinOE with the theory of symmetric
functions.

Certainly, more work is needed to accomplish the spectral theory of
GinOE. (i) In particular, the large-$n$ behavior \cite{KA} of the
probability function $p_{n,k}$ should be examined when the number
$k$ of real eigenvalues does or does not scale with the matrix
dimension $n$. The large-$n$ formulae of this kind would facilitate
a comparison of our exact theory with existing experimental
\cite{KDI-2000} and numerical (see the second paper in Ref.
\cite{QCD}) data. (ii) The calculation of all $(r_1,r_2)$-point
correlation functions for GinOE as defined below (\ref{jpdf=1}) is
yet another important problem to tackle. We believe that progress in
this direction can be achieved through a proper extension of the
Pfaffian integration theorem (\ref{theorem-b}).

We thank T. Seligman for kind hospitality at the Centro
Internacional de Ciencias (CIC) in Cuernavaca, Mexico, where this
work was initiated. Clarifying correspondence with A. Borodin
(Caltech), V. B. Kuznetsov (Leeds) and G. Olshanski (IITP, Moscow)
is greatly appreciated. A part of this work was done during the
visits to Brunel University West London (E.K.) and University of
Warwick (E.K. and G.A.) supported by a BRIEF grant from Brunel
University. The research by E.K. was partially supported by the
Israel Science Foundation through the grant No 286/04.


\begin{references}
\bibitem{RP} Throughout the paper, we use a parameterisation
$(w_1,\cdots,w_n)=(\lambda_1,\cdots,\lambda_k; z_1,{\bar
z}_1,\cdots, z_\ell,{\bar z}_\ell)$.
\bibitem{R-FR} A partition $\blambda$ of the size $|\blambda|=\ell$ written in
    the frequency representation $\blambda = (\ell_1^{\sigma_1},\cdots,
    \ell_g^{\sigma_g})$ implies that the part $\ell_j$ appears
    $\sigma_j$ times so that $\ell = \sum_{j=1}^g \ell_j\, \sigma_j$,
    where $g$ is the number of inequivalent parts of partition.
\bibitem{K-rem} Strictly speaking, the formulae (\ref{Dker}) and
(\ref{mu}) as they stand are only valid for $n=2m$ even. Since for
$n=2m+1$ odd a similar representation is available
\cite{M-2004,NN-2001} for the kernel $K_n(x,y)$, our derivation
holds for $n$ of any parity.
\end{references}
\end{document}